\documentclass[prb,twocolumn,showpacs,superscriptaddress]{revtex4}
\usepackage{amsmath}
\usepackage[hypertex]{hyperref}

\newcommand{\GL}{\mathrm{\scriptscriptstyle GL}}

\newcommand{\sigmaD}{\sigma_{\scriptscriptstyle D}}
\newcommand{\DOS}{\mathrm{\scriptscriptstyle DOS}}
\newcommand{\MT}{\mathrm{\scriptscriptstyle MT}}
\newcommand{\AL}{\mathrm{\scriptscriptstyle AL}}

\newcommand{\Rel}{\mathrm{Re}}

\newcommand{\Imag}{\mathrm{Im}}
\newcommand{\Real}{\mathrm{Re}}

\newcount\minute
\newcount\hour
\newcount\hourMins
\def\now
{
  \minute=\time
  \hour=\time \divide \hour by 60
  \hourMins=\hour \multiply\hourMins by 60
  \advance\minute by -\hourMins
  \zeroPadTwo{\the\hour}:\zeroPadTwo{\the\minute}
}\def\timestamp
{
  \today\ \now
}
\def\today
{
  \zeroPadTwo{\the\day}-\zeroPadTwo{\the\month}-\the\year%
}
\def\zeroPadTwo#1%
{
  \ifnum #1<10 0\fi
  #1%
}

\newcommand{\bnabla}{{\boldsymbol{\nabla}}} \newcommand{\Tr}{\mathrm{Tr}} \newcommand{\Dk}{\check{\Delta}_{\cal K}} \newcommand{\Qk}{\check{Q}_{\cal K}} 
 \newcommand{\Ak}{\check{\mathbf{A}}_{\cal K}} \newcommand{\Si}{\check{\Xi}}
\newcommand{\cK}{\cal K}

\newcommand{\bx}{{\mathbf x}}
\newcommand{\br}{{\mathbf r}}

\newcommand{\bq}{{\mathbf q}}

\newcommand{\dif}{{\mathrm d}}

\def\XXint#1#2#3{{\setbox0=\hbox{$#1{#2#3}{\int}$}
\vcenter{\hbox{$#2#3$}}\kern-.5\wd0}}

\date{\timestamp}

\begin{document}

\title{Pairing Fluctuation AC Conductivity of Disordered Thin Films}
\author{Aleksandra Petkovi\'{c}}
\affiliation{Laboratoire de Physique Th\'{e}orique et Hautes Energies, Universit\'{e} Pierre et Marie Curie and CNRS, 4 place Jussieu, 75005 Paris,  France
}
\author{Valerii M. Vinokur}
\affiliation{Materials Science Division, Argonne National Laboratory, Argonne, Illinois 60439, USA}

\begin{abstract}
We study temperature $T$ and frequency $\omega$ dependence of the in-plane fluctuation conductivity of a disordered superconducting film above the critical temperature. Our calculation is based on the nonlinear sigma model within the Keldysh technique. The fluctuation contributions of different physical origin are found and analyzed in a wide frequency range. In the low-frequency range, $\omega\ll T$, we reproduce the known leading terms and find additional subleading ones in the Aslamazov-Larkin and the Maki-Thompson contributions to the ac conductivity. We also calculate the density of states ac correction. In the dc case these contributions logarithmically depend on the Ginzburg-Landau rate and are considerably smaller that the leading ones.  However, in the ac case an external finite-frequency electromagnetic field strongly suppresses the known Alsamazov-Larkin and Maki-Thompson ac contributions, while the corresponding new terms and the density of states contribution are weakly suppressed and therefore become relevant at finite frequencies. \end{abstract}
\maketitle

\section{Introduction}

The behavior of conductivity of a superconductor as function of temperature in the vicinity of the superconducting transition is governed by superconducting fluctuations and implicitly contains microscopic and macroscopic parameters characterizing
the material.  This motivated the invariable importance of fluctuations as one of the main tools of superconducting studies\cite{Varlamovbook}.
One can identify three major processes contributing to fluctuation conductivity. Fluctuation-induced Cooper pairs, having the finite lifetime in the normal domains give rise to the increase of conductivity; this is the so-called Aslamazov-Larkin (AL) contribution\cite{Aslamazov1}. More involved contribution arises from the Andreev reflection of quasiparticles from superconducting fluctuations accounting for the interference
due to quasi-particle scattering on impurities; this is so-called Maki-Thompson (MT) contribution\cite{Maki,Thompson}.
The third fluctuation contribution to conductivity, the negative one, is due to the suppression of the quasi-particle density of states (DOS)
at the Fermi level \cite{Varlamovbook}.
Impact of all these processes on the dc conductivity has been investigated in detail close to the transition and far-from it \cite{Varlamovbook}.
Recently, the extensive studies of superconductor-insulator transition \cite{scinsul} and quest for the explanation of mechanisms of
high temperature superconductivity have revitalized the interest to physics of fluctuations\cite{Kikuchi,Tatjana}, and enriched
our understanding of the dc fluctuation conductivity\cite{Glatz1+,Finkelstein}.
In this work motivated by recent progress in experimental studies of the ac properties of disordered
superconductors\cite{exp2,exp1, exp3, exp4, exp5},
we address both temperature and frequency dependence of the fluctuation in-plane ac conductivity of films in a wide frequency and temperature range.
The calculations are carried out in the framework of the nonlinear sigma model for disordered superconductors
via employing the Keldysh technique\cite{Feigelman+00,KamenevAndrev}.
This approach describes the low-energy physics, and is valid at energy scales smaller that the quasiparticle elastic scattering rate
and appears adequate for the analysis of the fluctuation transport\cite{AlexPRB,NikolayEPL,PRL,PRBmy}.

The ac in-plane conductivity of superconducting films was studied in Refs.~\onlinecite{Schmidt,acMT}
using Matsubara diagrammatic technique and the low-frequency, $\omega\ll T$, in-plane AL and MT contributions were obtained in the leading order.
The approach chosen in our work has an advantage of being more physically transparent than the more traditional
Matsubara technique since it is formulated in the real-time representation and as such does not involve the analytic continuation.
Moreover it applies naturally to out-of-equilibrium situations\cite{NikolayEPL,PRL,PRBmy}.
Using Keldysh technique we obtain the ac conductivity in a wide frequency and temperature range. We re-derive the well-known leading low-frequency AL and MT contributions\cite{Schmidt,acMT} which are characterized by the power law dependence on the Ginzburg-Landau relaxation time in the dc limit and obtain the additional, subleading AL and MT contributions to the ac conductivity. In the dc limit these subleading corrections behave as the DOS contribution, i.e., logarithmically depend on the Ginzburg-Landau time. These subleading terms, being of little practical importance in the dc limit, become relevant in the ac case, since the leading AL and MT terms are strongly suppressed by the applied finite-frequency field, while the subleading ones are only weakly suppressed. We also calculate the DOS contribution to the ac conductivity and find the similar scenario. Although in the dc case the DOS contribution is subleading, it depends weakly on the external-field frequency $\omega$ and at intermediate frequencies $\omega\sim T$ it becomes of a similar order as the AL and the MT contributions.

Recent studies that have been addressing the behavior of dc fluctuation conductivity appear to diverge
in details of the final results (in the subleading terms).
From this viewpoint an alternative approach based on the
Keldysh technique is a step towards the understanding of the possible origin of the existing discrepances.
In the zero magnetic field, the subleading terms add up to
$\mathcal{C}  \zeta(3)e^2\ln{\left(T\tau_{\GL}\right)}/(\pi^4 d_f)$,
where $d_f$ is the film thickness, $\tau_{\GL}$ denotes the Ginzburg-Landau time
and $\mathcal{C}$ is a numerical coefficient.
The recent paper \cite{Glatz1+} finds the result $\mathcal{C}=-231/8$
using Matsubara diagrammatic technique.
A different result with $\mathcal{C}=-21$, is reported in another recent study\cite{Finkelstein}.
There, the calculation was based on the Usadel equation and it is very closely related to the
nonlinear sigma model approach.
The results of Refs.~\onlinecite{Glatz1+} and~\onlinecite{Finkelstein} for the individual
subleading fluctuation contributions agree only on the DOS correction.
As a special case of our results, at zero frequency we obtain $\mathcal{C}=-21/2$ and
an additional term depending logarithmically on the ratio between the phase coherence rate and the temperature.
The comparison and discussion of the expressions for the individual fluctuation contributions
is given in detail in Sec.~\ref{sec:discussion}.

Apart from the two mentioned papers, there are many other older studies of the dc conductivity,
based on the Matsubara technique, which do not include all the subleading terms.
We do not plan on involving ourselves into a comprehensive review of all the interesting works
and the detailed comparison of all the subtleties
since the classification of diagrams yielding fluctuation contributions to the conductivity in
Matsubara technique differs from the one within the approach based on the nonlinear sigma model.
Thus only when all the contributions are summed up, the results must coincide and can be compared.
We discuss these issues in more details below, see Sec.~\ref{sec:discussion}.

The paper is organized as follows. In Sec.~\ref{sec:model} we introduce the model and basic equations.
In Sec.~\ref{sec:MT}, the MT fluctuation contribution to the in-plane conductivity is analysed.
In Secs.~\ref{sec:AL} and \ref{sec:dos} we consider the AL and DOS contributions, respectively.
In Sec.~\ref{sec:discussion}, we discuss different methods and compare our findings with the results known in literature.
The summary of results is given in Sec.~\ref{sec:summary}.
In Appendices \ref{App:MT}, \ref{App:AL}, and \ref{App:DOS} we present expressions for MT, AL and DOS contributions to the conductivity for arbitrary frequency and Ginzburg-Landau time, respectively. In Appendix \ref{app:ALnoise}, we provide the calculation of the AL contribution by considering the current-current correlation function and using the fluctuation-dissipation theorem.

\section{Model and basic equations\label{sec:model}}

In this section we introduce the model and basic equations that will be used in the rest of the paper. We consider a disordered superconductor described by the Hamiltonian $H=H_0+H_{\rm int}$. The single-particle Hamiltonian in the coherent state basis reads as
\begin{align}
H_0=\int \dif {\mathbf r}\;\bar\psi_\alpha\left[-\frac{(\bnabla - ie\mathbf{A})^2}{2m} +U_{\rm dis} +e \phi \right]\psi_\alpha.
\end{align}
Hereafter $\hbar=c=k_B=1$. The fields $\mathbf A$, $\phi$ and $U_{\rm dis}$ are the vector, scalar and disorder potentials, respectively. The electron charge is denoted by $e$.  The spin variable is denoted by $\alpha\equiv\uparrow$,$\downarrow$, and the summation over the spin indices is implicitly assumed. The interaction is given by the Bardeen-Cooper-Schrieffer (BCS) Hamiltonian
\begin{align}
H_{\rm int}=
-\frac{\lambda}{\nu} \int\dif {\mathbf{r}} \; \bar\psi_{\uparrow}\bar\psi_\downarrow\psi_\downarrow \psi_\uparrow,
\end{align}
where the coupling constant $\lambda$ is positive. The disorder potential originates from quenched impurities and it is short-ranged. We assume that it is Gaussian distributed with the correlator
\begin{align}
\langle U_{\mathrm{dis}}(\br)U_{\mathrm{dis}}(\br')\rangle =\frac{1}{2\pi\nu\tau}\delta(\br-\br').
\end{align}
Here $\nu$ is the bare single particle density of states at the Fermi level per one spin projection and $\tau$ is the elastic scattering time.

In the following we use the Keldysh technique, that allows us to perform the disorder average without introducing replica fields. Then we carry out the standard decoupling of the four-fermion terms in the action via the Hubbard-Stratonovich fields $Q$ and $\Delta$, and after integrating out the degrees of freedom with the energies higher than the elastic scattering rate, we obtain the Keldysh nonlinear sigma model.\cite{L_Kamenev} The partition function takes the form \cite{Feigelman+00,KamenevAndrev}
\begin{align}
\label{eq:Z}
Z=\int \mathcal{D}Q\;\mathcal{D}\Delta\exp\{i S[\Qk,\Dk]\},
\end{align}
where the nonlinear sigma model action $S$ is the effective action describing the
low-energy physics at energy scales smaller than the elastic scattering rate $\tau^{-1}$.
It holds in the dirty limit where the motion of electrons forming fluctuation Cooper pairs is diffusive,
i.e., the lifetime of  Cooper pairs is much greater than the elastic scattering time.
The nonlinear sigma model action reads as
\begin{align}
\label{eq:S}
S[\Qk,\Dk]=S_{\Delta}+S_{\phi}+S_{Q},
\end{align}
where the different contributions are
\begin{align}
S_{\Delta}=&-\frac{\nu}{2\lambda}\Tr[\Dk^{\dagger}\check{Y}\Dk],\quad S_{\phi}=\frac{e^2\nu}{2}\Tr[\check \phi_{\mathcal{K}}\check{Y}\check \phi_{\mathcal{K}}],\\
\label{q:S_Q}
S_{Q}=&\frac{i \pi \nu}{4}\Tr[D (\partial_{\bf{r}}\Qk)^2-4 \Si\partial_t\Qk
-4ie\check{\phi}_{\mathcal{K}}\Qk\notag\\&+4i\Dk\Qk ].
\end{align}
The matrix field $\check{Q}$ satisfies $\check Q^2=1$.
Here $D=\tau v_F^2/d$ is the diffusion coefficient that carries information about the disorder, $v_F$ is the Fermi velocity, and $d$ denotes the system dimensionality. The check-mark $\check\ $ above the field variables indicates that they are defined in the space
that is the tensor product of the Keldysh and Nambu spaces.
The former and the latter are spanned by the Pauli matrices $\hat{\sigma}_i$ and $\hat{\tau}_i$, $i\in\{0,x,y,z\}$, respectively, and we define $\check{Y}=\hat{\sigma}_x\otimes \hat{\tau}_0$ and $\Si=\hat{\sigma}_0\otimes\hat{\tau}_z$. One uses different notation for the same matrices $\hat{\sigma}_i=\hat{\tau}_i$ for convenience, and $\hat{\sigma}_0=\mathrm{diag}(1,1)$. Multiplication in the time-space is implicitly assumed, and ``Tr'' includes an integration over real space. The subscript $\cK$ denotes the gauge transformed fields:
\begin{gather}
\check\phi_{\cK}=\check{\phi}-\partial_{t}\check{\cK},\\ {\Ak}={\check{\mathbf{A}}} +{\bf{\bnabla}}{\check{\cK}},\\ \check{\cK}=\left(k^{cl}\hat{\sigma}_{0}+k^{q}\hat{\sigma}_{x}\right)\otimes\hat{\tau}_{0}.
\end{gather}
The fields $\check A$ and  $\check \phi$ are defined in same way as $\check{\cK}$, while
\begin{gather}
\check{\Delta}= \left(\Delta^{cl}\hat{\sigma}_{0}+ \Delta^{q}\hat{\sigma}_{x}\right)\otimes\hat{\tau}_{+}-\mathrm{H.c.},\\
\Dk(\br,t)=e^{ie \Si{\check{\cK}}(\br,t)}\check{\Delta}e^{-ie\Si{\check{\cK}}(\br,t)}.
\end{gather}
The field $\Qk$ is defined in the same way. We have also defined $\hat{\tau}_{\pm}=(\hat{\tau}_x\pm i\hat{\tau}_y)/2$.
The quantum ($q$) and classical ($cl$) components of the fields are defined as the half-sum and the half-difference of the field values at the lower and the upper branches of the Keldysh time-contour, respectively. The field $\Delta^{cl}$ becomes the superconducting order parameter at the mean-field (saddle-point) level, while the saddle point equation for $\check Q$ produces the Usadel quasiclassical equations, where $\check Q$ plays the role of the quasiclassical Greens function. The covariant spatial derivative is given by
\begin{align}
\partial_{\mathbf{r}}\check{Q}_{\cK}= \bnabla_{\mathbf{r}}\check{Q}_{\cK}- ie[\check{\Xi}\check{\mathbf{A}}_{\cK},\check{Q}_{\cK}].
\end{align}

In the absence of the BCS interaction ($\lambda=0$), the metallic saddle point equation for $\check{Q}$ obtained from Eq.(\ref{eq:S}) reads as \cite{first,second,KamenevAndrev}
\begin{align}\label{eq:saddle}
\check \Lambda&=\check{\mathcal{U}}\check{\Lambda}_0 \check{\mathcal{U}}^{-1},\quad\quad\check{\Lambda}_0=\hat{\sigma}_z\otimes\hat{\tau}_z,\\
\label{lambda}
\check{\mathcal{U}}_{t,t'}(\br)&=\check{\mathcal{U}}_{t,t'}^{-1}(\br)=
\left(\begin{array}{cc}\delta(t-t'-0)&
F_{t,t'}(\br)\\
0&-\delta(t-t'+0)\end{array}\right)\otimes\hat{\tau}_0.
\end{align}
After Wigner transforming ${F}_{t,t'}(\br)$ we obtain $F_{\epsilon}(\br,t)$ which can be related to the  quasiparticle distribution function $f_{\epsilon}(\br,t)$ as $F_{\epsilon}(\br,t)= 1-2f_{\epsilon}(\br,t)$.
The strategy is to consider the massless fluctuations around the metallic saddle point solution. They can be conveniently parameterized as \cite{Feigelman+00}
\begin{align}\label{parametrization}
\check{Q}_{\cK}(\mathbf{r}) &= e^{- \check{W}(\mathbf{r})/2}\, \check{\Lambda}(\br)\,\, e^{\check{W}(\mathbf{r})/2 },\quad \check{W}=\check{\mathcal{U}}\check{\mathcal{W}} \check{\mathcal{U}}^{-1},\\
\label{eq:W}
\check{\mathcal{W}}&=\left(\begin{array}{cc}
w\tau_{+} - w^{*}\tau_{-} & w_{0}\tau_{0}+w_{z}\tau_{z}\\
\bar{w}_{0}\tau_{0}+\bar{w}_{z}\tau_{z} & \bar{w}\tau_{+} -
\bar{w}^{*}\tau_{-}
\end{array}\right),
\end{align}
such that $\check{W}\check{\Lambda}+\check{\Lambda}\check{W}=0$. Here we introduced four real fields $w^{\alpha}_{tt'}(\mathbf{r}),\bar{w}^{\alpha}_{tt'}(\mathbf{r})$
with $\alpha=0,z$ representing diffuson degrees of freedom and the two complex
fields $w_{tt'}(\mathbf{r}),\bar{w}_{tt'}(\mathbf{r})$ for Cooperon degrees of freedom. The asterisk denotes the complex conjugation.

In the following, we calculate in-plane ac conductivity of a thin disordered film in the normal state but in the close vicinity of the transition to the superconducting state, $T\gtrsim T_c$.  It can be found using the relation
\begin{align}\label{eq:acconductivity}
\sigma_{xx}(\omega)=-\frac{1}{2\omega}\int \dif \br \frac{\delta^2 Z}{\delta
\mathbf{A}_{{\cK},{x}}^{cl}(\br,\omega)\delta
\mathbf{A}_{{\cK},x}^{q}(\br,-\omega)}\Bigg|_{{\bf{A}}^q=0,{\bf{A}}^{cl}=0},
\end{align}
where $Z$ is the Keldysh partition function, Eq.~(\ref{eq:Z}). Here the index $x$ denotes the $x$-component of the vector field.
Since we are neither interested in the weak localization correction nor in the Altshuler-Aronov type corrections, but in contributions to the ac conductivity caused by fluctuations of the superconducting order parameter $\Delta(\br,t)$, in the following we consider only Cooperon degrees of freedom. We are interested in massless fluctuations around the metallic saddle point Eq.~(\ref{eq:saddle}) up to the second order in $\check{\mathcal{W}}$, Eq.~(\ref{eq:W}). This means that the film is not too close to the transition such that $T-T_c\geq G_i T_c$ where $G_i\ll 1$ is the Ginzburg number $G_i=(\nu D d_f)^{-1}$. Here $d_f$ denotes the film thickness. We consider linear response below, and then we can use the equilibrium distribution function $F_{\epsilon}(\br,t)=\tanh{\left(\epsilon/(2T)\right)}$ and calculate the correlation functions\cite{PRBmy},\footnote{Note that in Ref.~\onlinecite{PRBmy} a different notation is used, where $A(\bx)=\int\dif\bq A(\bq) \exp(i\bq\bx)/(2\pi)^d$ and $A^*(\bx)=\int\dif\bq A^*(\bq) \exp(i\bq\bx)/(2\pi)^d$ and similarly for the Fourier transform in the time-domain. Such notation is not convenient since it uses the convention that complex conjugation involves the change of momentum and energy, $\epsilon\to-\epsilon$ and $\bq\to-\bq$. In the following, we will rather use $A(\bx)=\int\dif\bq A(\bq) \exp(i\bq\bx)/(2\pi)^d$ and $A^*(\bx)=\int\dif\bq A^*(-\bq)\exp(i\bq\bx)/(2\pi)^d$. }:
\begin{widetext}
\begin{align}\label{corr1}
&\langle\langle
w_{\epsilon_1,\epsilon_2}({\bf{q}})w_{-\epsilon_3,-\epsilon_4}^*({\bf{q}})\rangle\rangle=\frac{2
i}{\nu}\delta_{\epsilon_1-\epsilon_2,\epsilon_4-\epsilon_3}\frac{-L^{-1}_{K,1-2}
L_{A,1-2}L_{R,1-2}+F_{\epsilon_3}L_{R,1-2}+F_{\epsilon_1}L_{A,1-2}}{\left[D
\mathbf{q}^2-i(\epsilon_1+\epsilon_2)\right]\left[D \mathbf{q}^2-i (\epsilon_3+\epsilon_4)\right]},\\ \label{corr2}
&\langle\langle
\bar{w}_{\epsilon_1,\epsilon_2}({\bf{q}})\bar{w}_{-\epsilon_3,-\epsilon_4}^*({\bf{q}})\rangle \rangle=\frac{2
i}{\nu}\delta_{\epsilon_1-\epsilon_2,\epsilon_4-\epsilon_3}\frac{-L^{-1}_{K,1-2}
L_{A,1-2}L_{R,1-2}-F_{\epsilon_2}L_{A,1-2}-F_{\epsilon_4}L_{R,1-2}}{\left[D
\mathbf{q}^2+i(\epsilon_1+\epsilon_2)\right]\left[D \mathbf{q}^2+i (\epsilon_3+\epsilon_4)\right]},
\\ \label{corr3}
&\langle\langle
\bar{w}_{\epsilon_1,\epsilon_2}({\bf{q}})w_{-\epsilon_3,-\epsilon_4}^*({\bf{q}})\rangle\rangle
= \frac{2
i}{\nu}\delta_{\epsilon_1-\epsilon_2,\epsilon_4-\epsilon_3}\frac{L^{-1}_{K,1-2}
L_{A,1-2}L_{R,1-2}+F_{\epsilon_2}L_{A,1-2}-F_{\epsilon_3}L_{R,1-2}}{\left[D
\mathbf{q}^2+i(\epsilon_1+\epsilon_2)\right]\left[D \mathbf{q}^2-i (\epsilon_3+\epsilon_4)\right]}.
\end{align}
\end{widetext}
Here the average $\langle\langle\ldots\rangle\rangle$ is with respect to the action $S$ given by Eq.~(\ref{eq:S}) and it includes averaging over the fluctuations of $\check{Q}$, $\Delta^{cl}$ and $\Delta^{q}$. Also,
$L_{R/A,i-j}\equiv\left(L_{R/A}^{-1}(\bq,\epsilon_i-\epsilon_j) \right)^{-1}$ denotes retarded/advanced fluctuation propagators and $L_{K,i-j}^{-1}\equiv L_{K}^{-1}(\bq,\epsilon_i-\epsilon_j)$ is the Keldysh propagator.
The fluctuation propagators read as\cite{AlexPRB,PRBmy}
\begin{gather}
L_R^{-1}(\mathbf{q},\omega)=\psi\left(\frac{1}{2}\right)- \psi\left(\frac{1}{2}+\frac{D\mathbf{q}^2-i\omega}{4\pi T}\right)-\frac{\pi}{8T\tau_{\scriptscriptstyle{GL}}},\\
L_A^{-1}(\mathbf{q},\omega)=[L_R^{-1}(\mathbf{q},\omega)]^*= L_R^{-1}(\mathbf{q},-\omega),\\
L_K^{-1}(\mathbf{q},\omega)=\coth\left(\frac{\omega}{2T}\right)
\left[L_R^{-1}(\mathbf{q},\omega)-L_A^{-1}(\mathbf{q},\omega)\right],
\end{gather}
where the Ginzburg-Landau rate is defined as
\begin{align}
\tau^{-1}_{\GL}=-\frac{8T}{\pi}\ln{\left(\frac{T_c}{T}\right)}.
\end{align}
Here we expressed the critical temperature as $T_c=2\omega_D e^{\gamma-1/\lambda}/\pi$, where $\omega_D$ is the Debye frequency and $\gamma$ is the Euler constant.

\section{Maki-Thompson ac conductivity\label{sec:MT}}

In this section we start the calculation of the ac conductivity using Eq.~(\ref{eq:acconductivity}). We remind the reader that all our results are valid for frequency $\omega$, and temperature $T$ much smaller than the elastic scattering rate $\tau^{-1}$, since this is the range of applicability of the nonlinear sigma model. Therefore in the absence of the BCS interaction $(\lambda=0)$, we obtain the Drude conductivity to be frequency independent:
\begin{align}\label{eq:Drude}
\sigmaD(\omega)&=-\frac{\pi \nu D e^2}{4\omega}\Tr\left(\check{\Lambda}_{\epsilon}\hat{\sigma}_x \check{\Lambda}_{\omega+\epsilon}+\hat{\sigma}_x\check{\Lambda}_{\epsilon} \check{\Lambda}_{-\omega+\epsilon} \right)\notag\\
&= \frac{\nu D e^2}{2\omega}\int\dif\epsilon\left( F_{\epsilon+\omega}-F_{\epsilon-\omega}\right)\notag \\
&=2\nu D e^2.
\end{align}
This result is valid for $\omega\tau\ll 1$. At higher frequencies, $\omega \tau\gg 1$, the real part od the Drude conductivity is expected to vanish as\cite{drude1,drude2} $\mathrm{Re}[\sigma_D(\omega)]\propto \omega^{-2}$, while the imaginary part behaves as $\mathrm{Im}[\sigma_D(\omega)]\propto\omega^{-1}$. The real part describes the dissipation in the system, i.e., the attenuation of the external electromagnetic field in the sample, while the imaginary part of conductivity gives information about its phase shift. If we know either real or imaginary part over a wide frequency range, the other one is determined by the Kramers-Kronig relation.

Next we include the BCS interaction and find the superconducting fluctuation corrections to the ac conductivity (\ref{eq:Drude}). In the introduction we already explained different processes that underlie different fluctuation contributions. We start with the Maki-Thompson contribution. Considering massless fluctuations around the metallic saddle point Eq.~(\ref{eq:saddle}) up to the second order in Cooperon degrees of freedom, and collecting all the terms of the type  $\langle\langle\bar{w}{w}^*\rangle\rangle$ and $\langle\langle{w}\bar{w}^*\rangle\rangle$ in Eq.~(\ref{eq:acconductivity}), we  find MT contribution:
\begin{widetext}
\begin{align}\label{eq:MT}
  \sigma_{\MT}(\omega)=&-\frac{\nu D e^2}{32 \pi^3}\frac{1}{\omega}\int \dif \epsilon_2\dif\epsilon_3 \dif\bq \left[F(\epsilon_2+\omega)-F(\epsilon_2)\right]\langle\langle \bar{w}_{\epsilon_2,\epsilon_3}(\bq) w^*_{-\epsilon_3-\omega ,-\epsilon_2-\omega}(\bq)+\bar{w}^*_{-\epsilon_2,-\epsilon_3}(\bq) w_{\epsilon_3+\omega ,\epsilon_2+\omega}(\bq)\rangle\rangle.
\end{align}
Here and in the following, $\int \dif\bq$ denotes the summation over the discrete component of the wave vector $\bq$ normal to the film ($q_z$) and integration over the continuous in-plane components ($q_x,q_y$). The former gives the factor $1/d_f$.

In the following we are interested in a system close to the transition to the superconducting state, such that  $\tau_{\GL}^{-1}\ll T$. Then, we find that the leading term reads as
\begin{align}\label{eq:MTleading}
\sigma_{\MT}(\omega)= -\frac{ie^2 D}{4\pi^2d_f}\frac{1}{\omega}\int_{-\infty}^{+\infty}\dif\epsilon_2\int_{-\infty}^{+\infty}\dif \epsilon \int_0^{\infty}q \dif q \frac{L^{-1}_{K}(\bq,\epsilon)|L_A(\bq,\epsilon)|^2\left[F(\epsilon_2+\omega)-F(\epsilon_2)\right]}
{(Dq^2+i(2\epsilon_2-\epsilon))(Dq^2-i(2\epsilon_2-\epsilon+2\omega))},
\end{align}
\end{widetext}
where $\omega>0$.
The main contribution in Eq.~(\ref{eq:MTleading}) comes from small momenta and frequency: $Dq^2,\epsilon\sim \tau_{\GL}^{-1}\ll T$. Therefore, the strategy is to carry out the expansion in this limit and obtain the series in powers of $T \tau_{\GL}$. For simplicity of presentation, in the following we consider two different limiting cases: i) $\tau_{\GL}^{-1},\omega\ll T$ and ii) $T,\omega\gg \tau_{\GL}^{-1}$. In order to obtain an expression valid for arbitrary frequency $\omega$, one has to consider also other contributions from Eq.~(\ref{eq:MT}), not contained in Eq.~(\ref{eq:MTleading}). We provide a complete expression valid at arbitrary $\tau_{\GL}$ and $\omega$ in Appendix \ref{App:MT}.

We start with the parameter region i).
Expanding the integrand in Eq.~(\ref{eq:MTleading}) for $D q^2,\omega,\epsilon \ll T$ we get the leading order terms
\begin{align}
\label{eq:MT11}
 \Rel[\sigma_{\MT}^{(1)}(x)]&= \frac{e^2}{2\pi d_f} {T}{\tau_{\GL}}\frac{\pi x-2\ln{\left(x\right)}}{1+x^2},\\
\label{eq:MT2}
\mathrm{Re}[\sigma_{\MT}^{(2)}(x)]=& -\frac{7 \zeta(3)e^2 }{\pi^4d_f}\Bigg[\ln{\left({T}{\tau_{\GL}}\right)}\notag\\ &+
   \frac{\pi  x^3
   +x^2-\left(3
   x^2+1\right) \ln
  \left (x\right)+1}{2\left(x^2+1\right)^2}\Bigg],
\end{align}
where $x=\omega\tau_{\GL}$. Indexes $(1)$ and $(2)$ correspond to the first and second largest contribution. The first term in Eq.~(\ref{eq:MT2}) is obtained with the logarithmic accuracy. Note that in Eqs.~(\ref{eq:MT11}) and (\ref{eq:MT2}) we assumed $\tau_{\GL}^{-1},\omega\gg D q^2_{\mathrm{min}}\approx\tau_{\phi}^{-1}$, where $\tau_{\phi}^{-1}$ denotes the phase breaking rate \cite{Varlamovbook}. In the case $\omega\ll \tau_{\phi}^{-1}\ll \tau_{\GL}^{-1}$, in Eqs.~(\ref{eq:MT11}) and (\ref{eq:MT2}) would appear $\tau_{\phi}^{-1}$ instead of $\omega$.
We point out that result given by Eq.~(\ref{eq:MT11}) as well as first term in Eq.~(\ref{eq:MT2}), are obtained using the lowest order expansion of the fluctuation propagator for $\epsilon,D q^2\ll T$:
$L_K^{-1}(\bq,\epsilon)=i{\pi}/{2}$ and $
L_{R/A}^{-1}(\bq,\epsilon)=
-\frac{\pi}{8T}(D\mathbf{q}^2+\tau_{\scriptscriptstyle{GL}}^{-1}\mp i\epsilon)$. In order to obtain the second summand in Eq.~(\ref{eq:MT2}) one has to use the next order expansion of the fluctuation propagators.

To summarize, the MT conductivity in the limit $\omega\ll T$ but for arbitrary ratio of $\omega$ and $\tau_{\GL}^{-1}$, is the sum
\begin{align}\label{eq:MTall}
\Rel[\sigma_{\MT}(x)]=&\Rel[\sigma_{\MT}^{(1)}(x)]
+\Rel[\sigma_{\MT}^{(2)}(x)],\quad\quad  x=\omega\tau_{\GL}.
\end{align}
The leading contribution is given by $\Rel[\sigma_{\MT}^{(1)}]$.
This is in agreement with Ref.~\onlinecite{acMT} where the MT contribution was studied in the
region i) and the first leading term $\Rel[\sigma_{\MT}^{(1)}(\omega)]$ was found.
In the dc case when $\tau_{\GL}^{-1}\gg \tau_{\phi}^{-1}$ we get the following dependence on the Ginzburg-Landau time from Eqs.~(\ref{eq:MT11}) and (\ref{eq:MT2}):
$
\Real[\sigma_{\MT}(\omega=0)]={e^2}{T}{\tau_{\GL}}\ln{(\tau_{\phi}/\tau_{\GL})}/({\pi d_f})-{7 \zeta(3)e^2 }\left[\ln{\left({T}{\tau_{\GL}}\right)}+
\ln{(\tau_{\phi}/\tau_{\GL})}/2\right]/({\pi^4d_f}).
$
We stress that the term proportional to $\ln{(\tau_{\phi}/\tau_{\GL})}$ originates from the last term in Eq.~(\ref{eq:MT2}) and in order to obtain it, on has to go beyond the lowest order term in the expansion of the propagators for small frequency and momentum.  Also, it is convenient to rewrite the result as
\begin{align}\label{eq:MTdc1}
\Real[\sigma_{\MT}(\omega=0)]=&\frac{e^2}{\pi d_f}{T}{\tau_{\GL}}\ln{(\tau_{\phi}/\tau_{\GL})} \notag\\&-\frac{7 \zeta(3)e^2 }{2\pi^4d_f}\left[\ln{\left({T}{\tau_{\GL}}\right)}+c
\ln{(T\tau_{\phi})}\right],
\end{align}
with a numerical coefficient $c$. From Eq.~(\ref{eq:MT2}) follows $c=1$. However, the previous analysis is focused on relevant dependence on $\tau_{\GL}$  that can be obtained using the expansion, and not on the dependence on $\tau_{\phi}$. Therefore we additionally analyzed complete MT expression (\ref{eq:MT}), both analytically and numerically. The complete analytical result is given in Appendix \ref{App:MT}. In the dc case it can be written as (\ref{eq:appMTdc}). We find that remaining contributions, not taken into account in Eqs.~(\ref{eq:MT11}) and (\ref{eq:MT2}) give contributions of the type $\ln{(T\tau_{\phi})}$ and change the numerical coefficient in front of this term from $c=1$ to $c\approx 3.5$. As expected, they do not influence singular dependence on $\tau_{\GL}$.

Now we consider the parameter region $\omega,T\gg \tau_{\GL}^{-1}$. The ratio of temperature $T$ and frequency $\omega$ is arbitrary. Then, in the similar way as in the case i), we obtain in the leading order two contributions
\begin{align}\label{eq:MT22}
\mathrm{Re}[\sigma_{\MT}^{(1)}(\omega)]=&\frac{4e^2}{\pi^2 d_f}\left(\frac{T}{\omega}\right)^2\ln{\left({T}{\tau_{\GL}}\right)}
\notag\\ &\times \mathrm{Re}\left[ \psi\left(\frac{1}{2}\right)-\psi\left(\frac{1}{2}-i\frac{\omega}{2\pi T}\right)\right],
\\\label{eq:MT112}
\mathrm{Re}[\sigma_{\MT}^{(2)}(\omega)]=&\frac{e^2 }{d_f}\left(\frac{T}{\omega}\right)^2\tanh\left(\frac{\omega}{2T}\right)
\end{align}
with the logarithmic accuracy. At intermediate and high frequencies $\omega\gtrsim T$, the terms $\sigma_{\MT}^{(1)}$ and $\sigma_{\MT}^{(2)}$ are of similar order with the weak domination of $\sigma_{\MT}^{(1)}$ as the frequency increases or the GL rate decreases, while for lower frequencies $\omega\ll T$ the leading contribution is given by Eq.~(\ref{eq:MT112}). The regions i) and ii) overleap at frequencies satisfying $\tau_{\GL}^{-1}\ll\omega\ll T$. Notice that there the subleading low-frequency term (\ref{eq:MT2}) matches the contribution (\ref{eq:MT22}), and also leading low-frequency contribution (\ref{eq:MT11}) matches the term (\ref{eq:MT112}). To conclude, analyzing regions i) and ii) we see that the leading term in the dc and low-frequency range, becomes more suppressed by the finite-frequency external electromagnetic field than the subleading term (in the low-frequency range), and therefore they become of similar magnitude at $\omega\gtrsim T$.

\section{Aslamazov-Larkin ac conductivity\label{sec:AL}}

In this section we calculate the Aslamazov-Larkin ac conductivity.  It is determined by the terms of Eq.~(\ref{eq:acconductivity}) which contain four fields that describe Cooperon degrees of freedom. We obtain
\begin{widetext}
\begin{align}\label{eq:AL}
\sigma_{\AL}(\omega)=&-\frac{(\pi\nu e D)^2}{2(2\pi)^8}\frac{1}{\omega}\int\dif \epsilon_1\dif\epsilon_2\dif\epsilon_4\dif\epsilon_5\dif\mathbf{q}_1 \dif\mathbf{q}_3 {q}_{1x} {q}_{3x}\notag\\
&
\Big\langle\Big\langle\big\{F(\epsilon_1) \left[\bar{w}^*(-\mathbf{q}_1,-\epsilon_2,-\epsilon_1+\omega) \bar{w}(-\mathbf{q}_1,\epsilon_1,\epsilon_2) -\bar{w}(\mathbf{q}_1,\epsilon_2,\epsilon_1-\omega) \bar{w}^*(\mathbf{q}_1,-\epsilon_1,-\epsilon_2)\right]\notag\\ &+F(\epsilon_1-\omega)\left[w(\mathbf{q}_1,\epsilon_2,\epsilon_1-\omega) w^*(\mathbf{q}_1,-\epsilon_1,-\epsilon_2) -w^*(-\mathbf{q}_1,-\epsilon_2,-\epsilon_1+\omega) w(-\mathbf{q}_1,\epsilon_1,\epsilon_2)\right]\big\}\notag\\
&[\bar{w}(\mathbf{q}_3,\epsilon_5,\epsilon_4+\omega)
\bar{w}^*(\mathbf{q}_3,-\epsilon_4,-\epsilon_5) -\bar{w}^*(-\mathbf{q}_3,-\epsilon_5,-\epsilon_4-\omega)
\bar{w}(-\mathbf{q}_3,\epsilon_4,\epsilon_5)\notag\\
&+ w(\mathbf{q}_3,\epsilon_5,\epsilon_4+\omega)
w^*(\mathbf{q}_3,-\epsilon_4,-\epsilon_5)- w^*(-\mathbf{q}_3,-\epsilon_5,-\epsilon_4-\omega)w(-\mathbf{q}_3,\epsilon_4,\epsilon_5)] \Big\rangle\Big\rangle
\end{align}
After performing Wick contractions, the summands of the previous expression can be classified with respect to the number of the distribution functions they contain: one, two, and  three. It turns out that the term that contains one distribution function $F$, exactly nullifies after the energy integration. The reason is that all the zeros of the corresponding denominators are in one half-plane of the complex plane. The term with two $F$ gives the leading contribution of Eq.~(\ref{eq:AL}) in the vicinity of the transition into superconducting state.  After exact integrations it reads as
\begin{align}\label{eq:AL1}
\sigma_{\AL}^{(1)}=&\frac{ie^2}{2\pi^2\alpha^3d_f} \int_{-\infty}^{+\infty}\dif y\int_0^\infty \dif x\frac{x\coth(2\pi y)}{|G(z)|^2G(z-i\alpha)}\Imag[G(z)][G(z)-G(z-2i\alpha)]\notag\\ &[G(z^*)+G(z-i\alpha)-G(z^*-i\alpha)-G(z-2i\alpha)].
\end{align}
\end{widetext}
Here we have introduced the short-hand notation
\begin{gather}
G(z)=\psi(1/2+z)-\psi(1/2)+\frac{\pi}{8T \tau_{\GL}},\\
\alpha=\frac{\omega}{4\pi T},\\
z=x+iy,\quad z^*=x-i y.
\end{gather}
Next we evaluate expression (\ref{eq:AL1}) analytically for small frequencies, while it can be evaluated numerically at arbitrary frequency. In the limit $\tau_{\GL}^{-1},\omega\ll T$ but for arbitrary ratio of $\omega$ and $\tau_{\GL}^{-1}$, we find
\begin{align}\label{eq:AL11}
\Real[\sigma_{\AL}^{(1,1)}]=&\frac{2e^2}{\pi d_f} (T\tau_{\GL})\frac{1}{\omega^2\tau_{\GL}^2} [\omega\tau_{\GL}\arctan(\omega\tau_{\GL}/2)\notag\\ &-\ln(1+\omega^2\tau_{\GL}^2/4)]\\\label{eq:AL12}
\Real[\sigma_{\AL}^{(1,2)}]=&\frac{7\zeta(3)e^2}{2\pi^4d_f}\frac{1}{\omega^2\tau_{\GL}^2} \Big[-4\omega\tau_{\GL}\arctan(\omega\tau_{\GL}/2)\notag\\ &+2\ln(1+\omega^2\tau_{\GL}^2/4) \notag\\ &-\omega^2\tau_{\GL}^2\ln\frac{1+\omega^2\tau_{\GL}^2/4}{T^2\tau_{\GL}^2}\Big]
\end{align}
The index $(1,1)$ denotes the leading term and $(1,2)$ the subleading term. Notice that in order to obtain the result (\ref{eq:AL12}) one has to go beyond the lowest order expansion of the fluctuation propagators for small frequency and momenta. The last term in Eq.~\ref{eq:AL12} is determined with the logarithmic accuracy.

Next we consider the term where the distribution function appears three times. Its full dependence for arbitrary $\tau_{\GL}$ and $\omega$ is given in Appendix \ref{App:AL}. Here we calculate its leading contribution for $\tau_{\GL}^{-1},\omega\ll T$. It is of the same order as $\sigma_{\AL}^{(1,2)}$, and reads as
\begin{align}\label{eq:AL21}
\sigma_{\AL}^{(2,1)}(\omega)=&-\frac{i e^2D^2}{\pi^2 d_f} \frac{1}{\omega}\Real[I_2]\Imag[I_1]\int_0^{\infty}\dif q q^3\int_{-\infty}^{+\infty}\dif \epsilon\notag\\&\times\{2 \Real[L_A(q,\epsilon) L_A(q,\epsilon-\omega)]\notag\\&+L_
R(q,\epsilon) L_A(q,\epsilon-\omega)\}.
\end{align}
In this equation $L_{A/R}$ denote the lowest order in the expansion of the fluctuation propagators for small frequencies and momenta. The functions $I_1$ and $I_{2}$ are defined as
\begin{align}
I_1=&-\int_{-\infty}^{+\infty}\dif\epsilon\frac{\tanh(\epsilon/2T)}{(2\epsilon-i0)^2}=
-i\frac{\pi}{8T},\\
\label{eq:I2}
I_2=&-\int_{-\infty}^{+\infty}\dif\epsilon\frac{\tanh^2(\epsilon/2T)}{(2\epsilon-i0)^2}=
-\frac{7\zeta(3)}{2\pi^2 T}.
\end{align}
After evaluation of expression (\ref{eq:AL21}), we obtain
\begin{align}\label{eq:realAL21}
\Real[\sigma_{\AL}^{(2,1)}]=&\frac{7\zeta(3)e^2}{2\pi^4d_f}\frac{1}{\omega\tau_{\GL}} \Bigg[-4\arctan(\omega\tau_{\GL}/2) \notag\\& -\omega\tau_{\GL}\ln\left(\frac{1+\omega^2\tau_{\GL}^2/4}{T^2\tau_{\GL}^2}\right)\Bigg],
\end{align}
with the logarithmic accuracy.

To summarize, in the limit $\tau_{\GL}^{-1},\omega\ll T$ we find the leading term $\sigma_{\AL}^{(1,1)}$ given by Eq.~(\ref{eq:AL11}) and the next leading order contribution is the sum of $\sigma_{\AL}^{(1,2)}$ and $\sigma_{\AL}^{(2,1)}$ given by Eqs.~(\ref{eq:AL12}) and (\ref{eq:realAL21}), respectively. They together give
\begin{widetext}
\begin{align}\label{eq:sigmaAL}
  \mathrm{Re}\left[\sigma_{\AL}(\omega)\right]=&\frac{2e^2}{\pi d_f}\frac{T}{\omega}\Bigg[
\arctan{\left(\frac{\omega \tau_{\GL}}{2}\right)}-
\frac{1}{\omega \tau_{\GL}}\ln{\left(1+\frac{\omega^2 \tau_{\GL}^{2}}{4}\right)}
\Bigg]\notag\\&+\frac{28 \zeta(3)e^2}{\pi^4 d_f}\frac{1}{\omega \tau_{\GL}}\Bigg[-\arctan{\left(\frac{\omega\tau_{\GL}}{2}\right)}
+\frac{1}{4 \omega\tau_{\GL}}
\ln\left(1+\frac{\omega^2\tau_{\GL}^{2}}{4}\right)+
\frac{\omega\tau_{\GL}}{4}\ln{\left(\frac{T^2\tau_{\GL}^{2}}{1+\omega^2\tau_{\GL}^{2}/4}\right)}
\Bigg].
\end{align}
\end{widetext}
The first leading term is in the agreement with calculation of Refs.~\onlinecite{Schmidt,acMT,acVarlamov}, while the second one is new. From Eq.~(\ref{eq:sigmaAL}), in the dc case we obtain
\begin{align}
\mathrm{Re}\left[\sigma_{\AL}(0)\right]=\frac{e^2}{2\pi d_f}{T}{\tau_{\GL}}+\frac{14\zeta(3)e^2}{\pi^4d_f}
\ln{\left({T}{\tau_{\GL}}\right)}.
\end{align}
Apart from the well known power law singularity of the AL dc contribution, we get the additional logarithmic divergence. Moreover, from Eq.~(\ref{eq:sigmaAL}), one gets that for $\omega\gg \tau_{\GL}^{-1}$, the subleading contribution is logarithmically suppressed with increasing $\omega$, while the leading term decays much faster as $\omega^{-1}$. So, although in the dc case the subleading term $\sigma_{\AL}^{(1,2)}+\sigma_{\AL}^{(2,1)}$ could be considered as negligibly small, in the ac case their sum becomes more relevant. In Appendix \ref{app:ALnoise} we rederive Eq.~(\ref{eq:sigmaAL}) calculating the current-current correlation function and using the fluctuation-dissipation theorem.

\section{Density of states ac conductivity\label{sec:dos}}

In this section we calculate the density of states contribution to the conductivity. We start form Eq.~(\ref{eq:acconductivity}) and collecting all the terms of the type $\langle \langle w w^*\rangle\rangle$ and $\langle \langle\bar{w}\bar{w}^*\rangle\rangle$ we  find
\begin{widetext}
\begin{align}\label{eq:generalDOS}
\sigma_{\DOS}(\omega)=&-\frac{\nu D e^2}{16 \pi^3}\frac{1}{\omega}\int\dif\epsilon_3\dif\epsilon_4 \dif\bq\Big\{F(\epsilon_3+\omega)\langle\langle \bar{w}_{\epsilon_3,\epsilon_4}(\bq)\bar{w}^*_{-\epsilon_4,-\epsilon_3}(\bq)\rangle\rangle-
F(\epsilon_3-\omega)\langle\langle {w}_{\epsilon_3,\epsilon_4}(\bq){w}^*_{-\epsilon_4,-\epsilon_3}(\bq)\rangle\rangle
\notag\\&-F(\epsilon_4) \langle\langle w_{\epsilon_3,\epsilon_4}(\bq)w^*_{-\epsilon_4-\omega,-\epsilon_3-\omega}(\bq)\rangle\rangle
+F(\epsilon_3)\langle\langle \bar{w}_{\epsilon_3,\epsilon_4}(\bq)\bar{w}^*_{-\epsilon_4+\omega,-\epsilon_3+\omega}(\bq)
\rangle\rangle
\Big\}
\end{align}
In Appendix \ref{App:DOS}, we provide an analytic form for arbitrary frequency and arbitrary Ginzburg-Landau time, while here we focus on the case when the system is close to the superconducting transition. Then, the leading contribution comes from the term
\begin{align}\label{eq:leadibgDOS}
\sigma_{\DOS}(\omega)=&+\frac{i D e^2}{4\pi^2d_f} \frac{1}{\omega} \int_{-\infty}^{+\infty}\dif\epsilon_3\dif\epsilon\int_0^{+\infty}\dif q q L^{-1}_K(\bq,\epsilon)|L_A(\bq,\epsilon)|^2 \Big\{
\frac{\tanh\left(\frac{\epsilon_3+\omega}{2T}\right)}{[D q^2+i(2\epsilon_3-\epsilon)]^2}-\frac{\tanh\left(\frac{\epsilon_3-\omega}{2T}\right)}{[D q^2-i(2\epsilon_3-\epsilon)]^2}\notag\\ & +\frac{\tanh\left(\frac{\epsilon_3}{2T}\right)}{[D q^2+i(2\epsilon_3-\epsilon)][D q^2+i(2\epsilon_3-\epsilon-2\omega)]}
-\frac{\tanh\left(\frac{\epsilon_3-\epsilon}{2T}\right)}{[D q^2-i(2\epsilon_3-\epsilon)][D q^2-i(2\epsilon_3-\epsilon+2\omega)]}\Big\}
\end{align}
We obtain the leading contribution for $\tau_{\GL}^{-1}\ll T,\omega$ to the ac conductivity to be
\begin{align}\label{eq:ac}
\mathrm{Re}\left[\sigma_{\DOS}(\omega)\right]=&\frac{2e^2 }{\pi^3d_f}\frac{T}{\omega}\ln{\left({T}{\tau_{\GL}}\right)}
\Bigg\{-\mathrm{Im}\left[\psi'\left( \frac{1}{2}-i\frac{\omega}{2\pi T}\right)\right]
+2\pi\frac{T}{\omega}\mathrm{Re}\left[\psi\left(\frac{1}{2}\right)-
\psi\left(\frac{1}{2}-i\frac{\omega}{2\pi T}\right)\right]\Bigg\}.
\end{align}
From this equation we get the limiting cases:
\begin{align}\label{eq:limitDOS}
\mathrm{Re}[\sigma_{\DOS}(\omega)]\approx&
\begin{cases} -\frac{21 e^2 \zeta(3)}{\pi^4d_f}\ln{\left({T}{\tau_{\GL}}\right)}, & \tau_{\GL}^{-1}\ll \omega\ll T\\
-\frac{4e^2}{\pi^2d_f} \ln{\left({T}{\tau_{\GL}}\right)}\left({T}/{\omega}\right)^2 \ln{\left({\omega}/{T}\right)},& \tau_{\GL}^{-1}\ll T\ll \omega. \end{cases}
\end{align}
\end{widetext}
Analyzing Eq.~(\ref{eq:generalDOS}) in the limit $\omega\ll \tau_{\GL}^{-1}\ll T$, we find
\begin{align}\label{eq:dosdc}
  \mathrm{Re}[\sigma_{\DOS}(\omega)]\approx-\frac{21 e^2 \zeta(3)}{\pi^4d_f}\ln{\left({T}{\tau_{\GL}}\right)}.
\end{align}
This result is also valid in the dc case, and we see that the DOS contribution is almost frequency independent at $\omega\ll T$.  Characteristic frequency for the DOS contribution is determined by temperature $T$. This is in contrast with the AL and the MT corrections whose characteristic frequency is determined by $T-T_c$. They decrease as power law $\sim \omega^{-1}$ with frequency, for $\tau_{\GL}^{-1}\ll\omega\ll T$. Therefore, at $\omega\sim T$ the DOS becomes of the same magnitude as the MT and the AL contributions.

Notice that the subleading contributions in other fluctuation corrections are of the same order as the leading DOS contribution at low frequencies. Also, notice that the DOS conductivity logarithmically  diverges for $\tau_{\GL}^{-1}\to 0$, even for finite frequency $\omega>0$, and it is negative. We remind the reader that all our results are valid for frequency, and temperature much smaller than the elastic scattering rate, since this is the range of applicability of the nonlinear sigma model.

\section{Discussions\label{sec:discussion}}

We are now equipped to compare different methods used for the analysis of the fluctuation transport
and discuss some of the results present in the literature.
First we would like to point out that classification of the diagrams into the different fluctuation contributions
to conductivity in the conventional Matsubara diagrammatic technique\cite{PRLVinokur,Varlamovbook} differs from that in
the approach based on the nonlinear sigma model, which we employ in our work.
In the latter, the physical origin of different contributions appears to be more transparent, as we discuss below.
However, when comparing our and the conventional diagrammatic approaches,
the final results summing up all the contributions in both calculations have to coincide.
More specifically, we expect that sum of our MT and DOS corrections should be equal to the
sum of the MT and DOS corrections in the diagrammatic approach, implying
that also the AL correction is the same in both approaches. For the classification of diagrams in the Matsubara diagrammatic approach see, e.g., Refs.~\cite{PRLVinokur,Varlamovbook}. 
The AL correction contains two contractions of the Cooperon fields and they correspond to
two fluctuation propagators appearing in the diagrams for the AL correction in the conventional approach.

Having said that, we turn to our classification of the fluctuation contributions.
All the terms in Eq.~(\ref{eq:acconductivity}) that contain a convolution of $\bar{w}$ and $w^*$
as well as $\bar{w}^*$ and $w$ give the MT contribution, while all the terms containing a convolution
of $w$ and $w^*$ as well as of $\bar{w}$ and $\bar{w}^*$ are collected into the DOS contribution to
the conductivity.
After redistribution of terms into the MT and the DOS contributions, the remaining ones belong to
the Alsamazov-Larkin contribution, Eq.~(\ref{eq:AL}).
These remaining terms contain four fields and applying the Wick theorem we get products of two contractions.

Let us analyze in more detail the structure of the terms in the DOS contribution.
We notice that the first and the second term in Eq.~(\ref{eq:generalDOS}) have slightly different structure than the third and fourth term.
Also, the expression for the change of the density of states of quasiparticles due to
superconducting fluctuations reads as
\begin{align}
\delta\nu(\epsilon)=-\frac{\nu}{32\pi^3}\int\dif\epsilon_1\int\dif \bq\langle\langle \bar{w}_{\epsilon,\epsilon_1}(\bq)\bar{w}^*_{-\epsilon_1,-\epsilon}(\bq) \notag\\+w_{\epsilon,\epsilon_1}(\bq)w^*_{-\epsilon_1,-\epsilon}(\bq) \rangle\rangle,
\end{align}
and then the first and the second term in Eq.~(\ref{eq:generalDOS}) in the dc case can be rewritten as
$2 D e^2\int \dif \epsilon \delta\nu (\epsilon) \partial_{\epsilon} F(\epsilon)$,
while the third and fourth term have different structure.
However, since all the contributions include the same type of the Cooperon degrees
of freedom, we group them into the DOS correction.

Note that the important progress in treating fluctuations within the Keldysh nonlinear sigma model was
achieved in Refs.~\onlinecite{AlexPRB,L_Kamenev}, where the leading fluctuation contributions
to the dc conductivity where  calculated.
However, there the contributions to the conductivity corresponding to the two last terms in
Eq.~(\ref{eq:generalDOS}) were missed. They obtained the factor $-7$ for the DOS correction. As follows from our consideration, in the dc case the first two summands in
Eq.~(\ref{eq:generalDOS}) give numerical
factor $-14$ and the last two give $-7$, and they together give factor $-21$ in Eq.~(\ref{eq:dosdc})
[note that there had been a misprint in the publication of the present authors, Ref.~\onlinecite{PRBmy},
in Eq.~(58), where the numerical factor $-21$ should stand instead of $-7$].

Next we summarize the results on the ac in-plane transport known in the literature
and compare them with our findings.
In Ref.~\onlinecite{Schmidt} the leading ac low-frequency AL term was found and later in
Ref.~\onlinecite{acMT} the leading low-frequency MT contribution was calculated.
These results are in an agreement with our results.
In Ref.~\onlinecite{acVarlamov} the ac DOS correction was obtained.
All these studies employed Matsubara  diagrammatic technique.
As we explained  above, the sum of the DOS and the MT contributions obtained in this approach
should be compared with our results.
However, because only the leading MT contribution was calculated in
Ref.~\onlinecite{acVarlamov} (it corresponds to the leading part of the so-called
anomalous contribution) and  the subleading MT terms that are of the similar magnitude as the
DOS contribution were not considered,
we can not compare our result for the DOS correction with the findings of Ref.~\onlinecite{acVarlamov}.

We now analyze the dc case where the subleading corrections in all the fluctuation contributions
were considered, contrary to the ac case, allowing us to undertake a more detailed comparison.
We have found in the previous sections that close to the superconducting transition
\begin{align}\label{eq:alldc}
  \Real[\sigma(\omega=0)]=&\frac{e^2}{2\pi d_f}T\tau_{\GL}\left[1+2\ln{\left(\tau_{\phi}/\tau_{\GL}\right)}\right]\notag\\&
  +\frac{e^2}{d_f}\frac{\zeta(3)}{\pi^4}\ln{\left(T\tau_{\GL}\right)}\left(c_{\AL}+c_{\MT}+c_{\DOS}\right)
  \notag\\&-\frac{e^2}{d_f}\frac{7\zeta(3)}{2\pi^4}c\ln{\left(T\tau_{\phi}\right)}
\end{align}
for $\tau_{\phi}\gg \tau_{\GL}$. Here $c_{i}$, where $i$ takes values $\AL,\MT, \DOS$ describes
contributions of different physical origin, the AL, MT and DOS. We find $c_{\AL}=14$, $c_{\MT}=-7/2$
and $c_{\DOS}=-21$.
The term $\ln{\left(T\tau_{\phi}\right)}$ comes from the MT contribution and $c\approx 3.5$.
The leading terms are given in the first line of Eq.~(\ref{eq:alldc}) and are well
established\cite{Varlamovbook}, unlike the remaining subleading terms.

The recent publication,  Ref.~\onlinecite{Finkelstein}, studied the dc conductivity using both the Usadel equation and the Matsubara technique. The results based on the Usadel equation give the total numerical coefficient $\mathcal{C}=\sum_i c_{i}=-21$ in front of the logaritmically
singular term $\ln{\left(T\tau_{\GL}\right)}$.
The authors of Ref.~\onlinecite{Finkelstein} used different gauge where the electric field is given by
$\bf{E}=-\bnabla \phi$,  and found  $c_{\MT}=-14$, $c_{\DOS}=-7$ and $c_{\AL}=0$.
In order to compare the results based on the Usadel equation of Ref.~\onlinecite{Finkelstein} with ours,
one has to consider the sum of the DOS and the AL terms, since the expressions for the MT dc contribution,
our Eq.~(\ref{eq:appMTdc}) and Eq.~(78) in Ref.~\onlinecite{Finkelstein}
for zero magnetic field, coincide\cite{Comparison}.
For the sum of the AL and the DOS terms there is full agreement, while the coefficients $c_{\MT}$ are different
due to inaccurate evaluation\cite{Comparison} of Eq.~(78) in Ref.~\onlinecite{Finkelstein}. The other approach in Ref.~\onlinecite{Finkelstein} based on the conventional diagrammatic approach gives the numerical coefficient $\mathcal{C}=-21$ as the sum of $c_{\DOS}=-14$, $c_{\AL}=+14$
and $c_{\MT}=-21$.
As we discussed at the beginning of this section, in order to compare our results with the ones based on
the conventional diagrammatic approach, one should compare the AL terms and the sum of the MT
and the DOS terms.
The result of Ref.~\onlinecite{Finkelstein} for the AL contribution agrees with our result.
However, there is a disagreement when comparing the sum of the subleading DOS
and  MT contributions of Ref.~\onlinecite{Finkelstein}.
Since Eq.~(\ref{eq:appMTdc}) also appears as a part of the MT contribution in the diagrammatic approach,
the above mentioned inaccurate evaluation of this term in Ref.~\onlinecite{Finkelstein} explains the
disagreement in the sum of the MT and the DOS terms. The two above comparisons of our findings with the results of Ref.~\onlinecite{Finkelstein} provides us with the comparison of each individual fluctuation contribution, and we infer agreement in the AL and the DOS results, while the disagreement in the MT term is resolved.

Note that in the recent paper of one of the present authors, Ref.~\onlinecite{Glatz1+}, it was found $\mathcal{C}=-231/8$ using the Matsubara diagrammatic technique.
(In Ref.~\onlinecite{Glatz1+} see Table 1 and region I.) There $c_{\AL}=-7/8$, $c_{\MT}=-14$,
and $c_{\DOS}=-14$.  The fact that the results on the subleading AL term of the present work agree with those
in Ref.~~\onlinecite{Finkelstein}, it adds to the confidence to validity of the latest findings that  $c_{\AL}=14$,
so the revealing the reasons for the remaining discrepancy in the subleading AL
contributions in Ref.~\onlinecite{Glatz1+} requires more analysis.
It also requires additional inspection if there should be an additional subleading MT contribution in
Ref.~\onlinecite{Glatz1+} that depends logarithmically on the ratio of the phase coherence time
 and the Ginzburg-Landau time that would change the coefficient $c_{\MT}=-14$ to $c_{\MT}=-21/2$,
 see discussion in Sec.~\ref{sec:MT} below Eq.~(\ref{eq:MTall}).

\section{Summary of results\label{sec:summary}}

We considered a disordered film above the superconducting transition
and studied influence of superconducting fluctuations on the ac in-plane conductivity.
Our approach is based on the nonlinear sigma model for
dirty superconductors that is valid for frequencies smaller than the elastic scattering rate.
Within this limitation, we provide analytical expressions for the ac in-plane conductivity for
arbitrary frequency $\omega$ and arbitrary Ginzburg-Landau time $\tau_{\GL}$.
We analyzed them in more detail in the following frequency regions: i) $\tau_{\GL}^{-1},\omega\ll T$
and arbitrary ratio between $\omega$ and $\tau_{\GL}^{-1}$,
and ii) $T,\omega\gg \tau_{\GL}^{-1}$ but arbitrary ratio between $\omega$ and $T$.
These regions have an overlap and provide us with conductivity behavior in the wide frequency range. We point out that possible additional effects due to nonthermal quasi-particle distribution, in the limit of high external frequencies, are beyond the scope of this work.

We calculated: (a) the Maki-Thompson contributions to the ac conductivity given by
Eqs.~(\ref{eq:MT11},\ref{eq:MT2}) in the region i) and by Eqs.~(\ref{eq:MT22},\ref{eq:MT112}) in the region ii);
 (b) the Aslamazov-Larkin contribution, Eq.~(\ref{eq:sigmaAL});
 (c) the density of states contribution given by Eq.~(\ref{eq:dosdc}) in the region i)
 and by Eq.~(\ref{eq:ac}) in the region ii).
 Our results agree with the existing leading MT and AL contributions to the conductivity,
 that were known for low frequencies $\omega\ll T$,  see Eqs.~(\ref{eq:MT11}) and (\ref{eq:AL11}).
The obtained expressions for the subleading low-frequency behavior of the AL  [second line of Eq.~(\ref{eq:sigmaAL})] and the MT [Eq.~(\ref{eq:MT2})] corrections,
 and intermediate- and high-frequency behavior of the MT [Eqs.~(\ref{eq:MT22}) and (\ref{eq:MT112})] and the DOS [Eq.~\ref{eq:ac}] corrections  are new and provide us with a complete physical picture.  For example, in the dc case the DOS contribution could be ignored with respect to the others.
 However, at frequencies $\omega\ll T$ it is almost frequency-independent in contrast to the other two contributions which get strongly suppressed by the external electromagnetic field.
 Therefore the DOS contribution becomes of the same magnitude as compared to the other two ones for $\omega\sim T$.

A similar approach to the one employed here can be used for calculation of the c-axis transport.
In the measurements of the c-axis reflectivity spectra in $\mathrm{YBa_2Cu_4O_8}$ single crystals
in Refs.~\onlinecite{acexperiment1},\onlinecite{acexperiment2} it was found
that the c-axis optical conductivity shows a transition from a Drude-like to a pseudo-gap like behavior
with the decrease of the temperature.
In Ref.~\onlinecite{acVarlamov}, the influence of the superconducting fluctuations on the c-axis conductivity
was considered and the non-monotonic frequency behavior that corresponds to the pseudo-gap was found.
Its origin is related to the nonmonotonic frequency behavior of the DOS correction.
Note that in the case of c-axis transport, the positive AL and MT contributions are suppressed by the
interlayer transmittance and compete with the negative DOS contribution\cite{Ioffe+,acVarlamov}.
At the intermediate frequencies, the DOS correction dominates and one obtains nonmonotonic
behavior\cite{acVarlamov}.
However, the difference between the minimal and the maximal conductivity of this nonmonotonic behavior
is small and only logaritmically increases as the transition is approached.
Therefore, it would be interesting to calculate subleading AL and MT contributions and see how they
influence this result. In the c-axis conductivity the leading AL term behaves as $\tau_{\GL}^2$
and the leading MT term as $\tau_{\GL}$, while the DOS term shows logarithmic dependence on the
Ginzburg-Landau time \cite{acVarlamov}.
One can expect that the subleading term in the AL contribution might depend linearly on
the Ginzburg-Landau time and might play a role in the final result as the transition is approached.

\section{Acknowledgments}
The authors are grateful to Z.~Ristivojevic for numerous helpful remarks. We thank K.~S.~Tikhonov and A.~A.~Varlamov for useful discussions. A.~P.~acknowledges the support from the ANR Grant No. ANR-2011-BS04-012. V. M. V. work is supported by the U.S. Department of Energy, Office of Basic Energy Sciences under contract no. DE-AC02-06CH11357. 
\appendix

\section{Maki-Thompson contribution\label{App:MT}}

In this appendix we present the complete expression for the MT contribution (\ref{eq:MT}), valid for arbitrary $\tau_{\GL}$ and arbitrary frequency. After performing one integration, it can be written as the sum of the following terms:
\begin{widetext}
\begin{align}\label{eq:MTallAPP1}
\sigma_{\MT1}&=\frac{-i}{8\pi^2\alpha} \int_{-\infty}^{+\infty}\dif y\int_0^\infty \dif x \frac{\coth(2\pi y)}{x-i\alpha}\frac{\Imag[G(z)]}{|G(z)|^2}\left\{2\Real[G(z)]-G(z-2i\alpha)-G(z^*-2i\alpha)\right\},\\
\label{eq:MTallAPP2}
\sigma_{\MT2}=&\frac{1}{8\pi^2\alpha} \int_{-\infty}^{+\infty}\dif y\int_0^\infty \dif x\big\{-i\pi+\coth(2\pi y)[2i\Imag[G(z)]+G(z-2i\alpha)-G(z^*-2i\alpha)]\notag\\
&-\coth[2\pi(y+\alpha)][2G(z)-2i \Imag[G(2i(y+\alpha))]-2G(z^*-2i\alpha)]\big\} \frac{1}{(x-i\alpha)G(z)}.
\end{align}
Here we have introduced the short-hand notation
\begin{gather}\label{G}
G(z)=\psi(1/2+z)-\psi(1/2)+\frac{\pi}{8T \tau_{\GL}},\\\label{alpha}
\alpha=\frac{\omega}{4\pi T},\\
z=x+iy,\quad z^*=x-i y\label{z}.
\end{gather}
We used the integral
\begin{align}\label{eq:integral}
\int_{-\infty}^{+\infty}\dif x\frac{\tanh x}{(x+a)(x+b)}= \frac{2}{a-b}\left[\psi\left(\frac{1}{2}-\frac{i a}{\pi}\right)- \psi\left(\frac{1}{2}-\frac{i b}{\pi}\right)\right]+
\begin{cases}
 0& \mathrm{Im}(a),\mathrm{Im}(b)>0,\\
\frac{2i \pi\tanh a}{a-b} & \mathrm{Im}(a)<0,\mathrm{Im}(b)>0
\end{cases},
\end{align}
and the identity
\begin{align}
  \psi^{(n)}(1-z)=(-1)^n \psi^{(n)}(z)+(-1)^n\pi \frac{\partial^n \cot{\pi z}}{\partial z^n},
\end{align}
for a positive integer number $n$.
Note that $\sigma_{\MT1}$ coincides with Eq.~(\ref{eq:MTleading}), while the remaining terms in Eq.~(\ref{eq:MT}) are denoted by $\sigma_{\MT2}$. The main contribution close to the superconducting transition comes from $\sigma_{\MT1}$, as discussed in section \ref{sec:MT}.

In the dc case from Eqs.~(\ref{eq:MTallAPP1}) and (\ref{eq:MTallAPP2}) we find
\begin{align}\label{eq:appMTdc}
\Real[\sigma_{\MT}(\omega=0)]=\int_{-\infty}^{+\infty}\dif y\int_{x_{\mathrm{min}}}^\infty \dif x
\frac{\Imag^2[G(z)]}{\pi x \sinh^2{(2\pi y)} |G(z)|^2}
\end{align}
Note that in the dc case one should regularize these integrals using the phase coherence time $\tau_{\phi}^{-1}$ which determines\cite{Varlamovbook} $x_{\mathrm{min}}\approx \tau_{\phi}^{-1}/(4\pi T)$ .

\section{AL contribution\label{App:AL}}

In this section we provide the complete expression for the AL contribution (\ref{eq:AL}) valid for arbitrary frequency and arbitrary Ginzburg-Landau time. It can be written as the sum of the two terms, one is given by Eq.~(\ref{eq:AL1}) and the other one is given by
\begin{align}
\sigma_{\AL}^{(2)}=&\int_{-\infty}^{+\infty}\dif y\int_0^\infty \dif x\Big\{-\coth (2 \pi  y-2 \pi  \alpha )\frac{x
   \left[G\left(z^*\right)-G\left(z^*
   -i \alpha \right)\right]
   \left\{G\left(z^*\right)+2i\Imag[G(z-i
   \alpha )]-G(z-2 i \alpha
   )\right\}}{4 \pi ^2 \alpha ^3
   G\left(z^*\right) G(z-i \alpha )}\notag\\+&
   \frac{x \coth (2 \pi  y)\left\{-G\left(z^*-i \alpha
   \right)+G(z-i
   \alpha )-2i\Imag[G(z)]\right\}
   }{8 \pi ^2
   \alpha ^3 |G(z)|^2
   G(z-i \alpha ) G\left(z^*-i \alpha
   \right)}
   \notag\\ &\times\Big(G\left(z^*\right)
   \left\{-G(z-2 i \alpha )
   G\left(z^*-i \alpha \right)+2 G(z)
   \left[G\left(z^*-i \alpha
   \right)+G(z-i \alpha
   )\right]-G(z-i \alpha
   )^2\right\}
   \notag\\&-G(z) \left[G\left(z^*-i
   \alpha \right)^2+G(z-i \alpha )
   G\left(z^*-2 i \alpha
   \right)\right]\Big)
   \notag\\&+
    \frac{x \coth (2 \pi  \alpha )}{8 \pi ^2
   \alpha ^3 |G(z)|^2
   G(z+i \alpha ) G\left(z^*+i \alpha
   \right)}
   \notag\\&\times \Big[G\left(z^*\right)
   \Big([G(z-i \alpha )-G(z+i \alpha
   )] [G(z-i \alpha )-G(z+i \alpha
   )+G(z+2 i \alpha )] G\left(z^*+i
   \alpha \right)\notag\\
   &-G(z) \left\{G(z+i
   \alpha ) \left[G\left(z^*-i \alpha
   \right)-2 G\left(z^*+i \alpha
   \right)\right]+|G(z-i \alpha )|^2
   \right\}\Big)\notag\\
   &+G(z) G(z+i
   \alpha ) \left[G\left(z^*+i \alpha
   \right)-G\left(z^*-i \alpha
   \right)\right] \left[-G\left(z^*-i
   \alpha \right)+G\left(z^*+i \alpha
   \right)-G\left(z^*+2 i \alpha
   \right)\right]\Big]\Big\}
\end{align}
The main contribution from this term when the system is close to the transition is given in Eq.~(\ref{eq:AL21}).

In the dc case we find that total AL contribution for arbitrary $\tau_{\GL}$ can be written as
\begin{align}
  \Real[\sigma_{\AL}](\omega=0)=&\int_{-\infty}^{+\infty}\dif y\int_0^\infty \dif x\Big\{
  \frac{x \Imag^2[G'[z]]}{\pi |G(z)|^2\sinh^2{(2\pi y)}}
  \notag\\&-\frac{x}{4\pi^3 |G(z)|^6}\left\{ |G(z)|^2\Real[G^2(z^*)G'(z)G'''(z)]-2\Real[G^3(z)[G'(z^*)]^2 G''(z^*)]\right\}\notag\\&
-\frac{x}{\pi^2}\coth{(2\pi y)} \Big\{ \frac{\Imag[G'(z)/G(z^*)]|G'(z)|^2+\Imag\{[G'(z^*)]^3/G(z^*)\}
+2\Imag[G'(z)G''(z)]}{2|G(z)|^2}
\notag\\ &
+\Imag\left\{\left[\frac{G'(z)}{G(z)}\right]^3\right\}-2\Imag\left[\frac{G'(z)G''(z)}{G^2(z)}\right]\Big\}
  \Big\}.
\end{align}
This expression contains many terms that are unimportant close to the transition, see Sec.~\ref{sec:AL}.

\section{DOS contribution\label{App:DOS}}

In this section we provide the complete expression for the DOS contribution (\ref{eq:generalDOS}), valid for arbitrary frequency and arbitrary Ginzburg-Landau time. It can be written as the sum of the two terms:
\begin{align}\label{DOS1}
\sigma_{\DOS1}=&\frac{1}{4\pi^2\alpha^2}\int_{-\infty}^{+\infty}\dif y\int_0^\infty \dif x\frac{\coth(2\pi y)}{|G(z)|^2}\Imag[G(z)]\left[G(z^*)-G(z^*-2i\alpha)+ i\alpha[G'(z-2i\alpha)+G'(z^*-2i\alpha)]\right],\\\label{DOS2}
\sigma_{\DOS2}=&\frac{i}{8\pi^2\alpha^2}\int_{-\infty}^{+\infty}\dif y\int_0^\infty \dif x\bigg\{\left[ \frac{2i\Imag[G(z)]+G(z^*-2i\alpha)-G(z-2i\alpha)}{G(z)}+2i\alpha \frac{G'(z-i\alpha)-G'(z^*-i\alpha)}{G(z-i\alpha)}\right]\coth(2\pi y)\notag\\&
-\frac{\coth(2\pi\alpha)}{G(z)}[2G(z)-G(z-2i\alpha)-G(z+2i\alpha)-2i\alpha G'(z)+2i\alpha G'(z-2i\alpha)]\bigg\},
\end{align}
we used the same notation as in the previous appendix, Eqs.~(\ref{G}),(\ref{alpha}), and (\ref{z}).
Note that here $\sigma_{\DOS1}$ coincides with Eq.~(\ref{eq:leadibgDOS}), while the remaining terms in Eq.~(\ref{eq:generalDOS}) are denoted by $\sigma_{\DOS2}$. The main contribution close to the superconducting transition comes from $\sigma_{\DOS1}$, as discussed in section \ref{sec:dos}.
In the dc case, from Eqs.~(\ref{DOS1}) and (\ref{DOS2}) follows
\begin{align}
\Real[\sigma_{\DOS}(\omega=0)]=\int_{-\infty}^{+\infty}\dif y\int_0^\infty \dif x \Big\{-\Real\left[\frac{G'''(z)}{4\pi^3G(z)}\right]+\frac{\coth{(2\pi y)}}{4\pi^2|G(z)|^2}\Big\{-4\Real[G(z)]\Imag[G''(z)]+6\Imag[G(z)]\Real[G''(z)]
\notag\\+\frac{2\Imag[G'(z)]\Real[G^2(z^*)G'(z)]}{|G(z)|^2} \Big\} \Big\}.
\end{align}
However, it turns out that many terms in this general expression are unimportant close to the superconducting transition, see Sec.~\ref{sec:dos}.
\end{widetext}

\section{AL conductivity\label{app:ALnoise}}

In this section we calculate the Aslamazov-Larkin ac conductivity in a different manner than in the main text. First, we calculate current-current correlation function $S(t-t',\br-\br')= \frac{1}{2}\langle\langle \mathbf{j}_x(\br,t)\mathbf{j}_x(\br',t')+\mathbf{j}_x(\br',t')\mathbf{j}_x(\br,t)\rangle\rangle$. It can be obtained as
\begin{align}\label{eq:noisedef}
S(t-t',\br-\br')=-\frac{1}{4}\frac{\partial^2 Z}{\partial {\bf{A}}^q_x(\br,t)\partial{\bf{A}}^q_x(\br',t')}\big|_{{\bf{A}}^q=0,{\bf{A}}^{cl}=0},
\end{align}
where $Z$ is the partition function defined in Sec.~(\ref{sec:model}).
Then, we use the fluctuation-dissipation theorem (FDT) and find real part of conductivity $\sigma$. FDT connects the current fluctuations and real part of conductivity $\sigma$ that tells us about absorbed energy in the sample
\begin{align}\label{eq:FDT}
S(\epsilon)=\mathrm{Re}[\sigma(\epsilon)] \epsilon \coth{\left(\frac{\epsilon}{2T}\right)}.
\end{align}
Here $S(\epsilon)$ is the Fourier transform of $S(t)=\int \dif\br S(t,\br)$. We find
\begin{widetext}
\begin{align}\label{eq:AL}
S_{\AL}(\epsilon)=&-\frac{1}{d_f}\left(\frac{\pi \nu e D}{2}\right)^2\int _{\epsilon_1,\epsilon_2,\epsilon_4,\epsilon_5,\bq_1,\bq_3} q_{1,x}q_{3,x}\Big\langle\Big\langle \Big[ -F(\epsilon_1) \bar{w}_{\epsilon_2,\epsilon_1-\epsilon}(\bq_1)
\bar{w}^*_{-\epsilon_1,-\epsilon_2}(\bq_1)\notag\\&+
F(\epsilon_1)\bar{w}^*_{-\epsilon_2,-\epsilon_1+\epsilon}(-\bq_1)
\bar{w}_{\epsilon_1,\epsilon_2}(-\bq_1)
+F(\epsilon_1-\epsilon) {w}_{\epsilon_2,\epsilon_1-\epsilon}(\bq_1)
{w}^*_{-\epsilon_1,-\epsilon_2}(\bq_1)\notag\\&-
F(\epsilon_1-\epsilon){w}^*_{-\epsilon_2,-\epsilon_1+\epsilon}(-\bq_1)
{w}_{\epsilon_1,\epsilon_2}(-\bq_1)
\Big]
\Big[ -F(\epsilon_4) \bar{w}_{\epsilon_5,\epsilon_4+\epsilon}(\bq_3)
\bar{w}^*_{-\epsilon_4,-\epsilon_5}(\bq_3)\notag\\&+
F(\epsilon_4)\bar{w}^*_{-\epsilon_5,-\epsilon_4-\epsilon}(-\bq_3)
\bar{w}_{\epsilon_4,\epsilon_5}(-\bq_3) +F(\epsilon_4+\epsilon) {w}_{\epsilon_5,\epsilon_4+\epsilon}(\bq_3)
{w}^*_{-\epsilon_4,-\epsilon_5}(\bq_3)\notag\\
&-
F(\epsilon_4+\epsilon){w}^*_{-\epsilon_5,-\epsilon_4-\epsilon}(-\bq_3)
{w}_{\epsilon_4,\epsilon_5}(-\bq_3)\Big]\Big\rangle\Big\rangle.
\end{align}
where $\int_{\epsilon_i}\equiv\int\dif\epsilon_i/(2\pi)$ and $\int_{\bq_i}\equiv\int\dif \bq_i/(2\pi)^2$.
The leading term close to transition to the superconducting state is
\begin{align}\label{eq:leadingAL}
  S_{\AL}^{(1)}=&-\frac{e^2 D^2}{8\pi^2}\int_0^{+\infty} \dif q q^3\int_{-\infty}^{+\infty} \dif \omega |L^A(\bq,\omega)|^2|L^A(\bq,\omega-\epsilon)|^2L^{-1}_K(\bq,\omega)L^{-1}_K(\bq,\omega-\epsilon)
  \notag\\
  &\times\left| \int_{-\infty}^{+\infty}\dif\epsilon_1\left\{\frac{\tanh{\left(\frac{\epsilon_1}{2T}\right)}}{\left[D q^2+i(2\epsilon_1+\omega-2\epsilon)\right]\left[D q^2+i(2\epsilon_1+\omega-\epsilon)\right]}+\frac{\tanh\left(\frac{\epsilon_1}{2T}\right)}{\left[D q^2+i(2\epsilon_1-\omega)\right]\left[D q^2+i(2\epsilon_1-\omega-\epsilon)\right]}\right\}\right|^2.
\end{align}
Using
\begin{align}\label{eq:integral1}
\int_{-\infty}^{+\infty}\dif x\frac{\tanh x}{(x+a)(x+b)}= \frac{2}{a-b}\left[\psi\left(\frac{1}{2}+\frac{i a}{\pi}\right)- \psi\left(\frac{1}{2}+\frac{i b}{\pi}\right)\right], \quad \mathrm{Im}(a),\mathrm{Im}(b)<0,\\
\end{align}
we perform the integration over $\epsilon_1$. Then, analyzing the obtained expression, we find that the main contribution comes from small momenta and for frequency $|\omega|\lesssim \tau_{\GL}^{-1}$ and $|\omega-\epsilon|\lesssim \tau_{\GL}^{-1}$.  In the limit $\tau_{\GL}^{-1},\epsilon\ll T$ but for an arbitrary ratio of $\epsilon$ and $\tau_{\GL}^{-1}$, we find
\begin{gather}\label{eq:SAL11}
S_{\AL}^{(1,1)}(\epsilon)=\frac{4 e^2}{\pi d_f}\frac{ T^2}{\epsilon^2}\tau_{\GL}^{-1}{ \left[-\ln\left(1+\frac{\epsilon^2\tau_{\GL}^{2}}{4}\right)+\tau_{\GL}  \epsilon  \arctan\left(\frac{\epsilon\tau_{\GL}
}{2}\right)\right]},
\\\label{eq:SAL12}
S_{\AL}^{(1,2)}(\epsilon)= \frac{14 e^2 \zeta (3)}{\pi ^4
d_f}\frac{T}{\epsilon ^2\tau_{\GL}^2}\left[4\epsilon \tau_{\GL}\arctan\left(\frac{\epsilon\tau_{\GL}}{2}\right) +\ln\left(1+\frac{\epsilon^2\tau_{\GL}^{2}}{4}\right) -\epsilon^2 \tau_{\GL}^2 \ln\left(\frac{4T^2 \tau_{\GL}^{2}}{4+\epsilon^2 \tau_{\GL}^{2}}\right)\right].
\end{gather}
\end{widetext}
Here the first line gives the leading and the second one the subleading  contribution,  $S_{\AL}^{(1)}(\epsilon)\approx S_{\AL}^{(1,1)}(\epsilon)+S_{\AL}^{(1,2)}(\epsilon)$.

We calculate further an additional contribution to the AL noise, originating from Eq.~(\ref{eq:AL}), and not taken into account in Eq.~(\ref{eq:leadingAL}). It is also singular function for $\tau_{\GL}^{-1}\to 0$ in the dc limit, but less relevant than the expression (\ref{eq:SAL11}). It gives a contribution to the noise that is of the same order as the subleading term (\ref{eq:SAL12}). The Aslamazov-Larkin noise, Eq.~(\ref{eq:AL}), contains products of two correlation functions given by Eqs.~(\ref{corr1}-\ref{corr3}). In  the leading contribution (\ref{eq:leadingAL}), only the terms $\sim L^{-1}_K$ were taken into account from the correlators (\ref{corr1}-\ref{corr3}). In the present case, when calculating product of the correlation functions, we should take into account the product of one term $\sim L^{-1}_K$ and one that contains only a retarded or advanced fluctuation propagator.  Then we find the leading term
\begin{align}\label{eq:subleadingAL}
S_{\AL}^{(2,1)}(\epsilon)=&\frac{i e^2D^2}{\pi^2 d_f}\mathrm{Im}(I_1)
\mathrm{Re}(I_2)
\notag\\&\times\int_{0}^{+\infty} \dif q q^3\int_{-\infty}^{+\infty} \dif\omega |L_A(q,\omega)|^2 L_K^{-1}(q,\omega)\notag\\&\times\mathrm{Re}\left[L_A(q,\omega-\epsilon)+L_A(q,\omega+\epsilon)\right]
.
\end{align}
Here, only the lowest order term in the expansion of the fluctuation propagators for small $\omega$ and $q$ appears. The functions $I_1$ and $I_{2}$ are defined as
\begin{align}
I_1=&-\int_{-\infty}^{+\infty}\dif\epsilon\frac{\tanh(\epsilon/2T)}{(2\epsilon-i0)^2}=
-i\frac{\pi}{8T},\\
\label{eq:I2}
I_2=&-\int_{-\infty}^{+\infty}\dif\epsilon\frac{\tanh^2(\epsilon/2T)}{(2\epsilon-i0)^2}=
-\frac{7\zeta(3)}{2\pi^2 T}.
\end{align}
Evaluating the expression (\ref{eq:subleadingAL}) we obtain
\begin{align}\label{eq:SAL21}
S_{\AL}^{(2,1)}=&-\frac{112 \zeta(3)e^2}{\pi^4 d_f}T\frac{\tau_{\GL}^{-1}}{\epsilon}\Bigg[\arctan{\left(\frac{\epsilon}{2\tau_{\GL}^{-1}}\right)}
\notag\\&-\frac{\epsilon}{4\tau_{\GL}^{-1}}\ln{\left(\frac{T^2}{\tau_{\GL}^{-2}+\epsilon^2/4}\right)}
\Bigg].
\end{align}
Therefore, in the limit $\tau_{\GL}^{-1},\epsilon\ll T$ we obtain the leading term $S_{\AL}^{(1,1)}$ given by Eq.~(\ref{eq:SAL11}) and the next leading order is the sum of $S_{\AL}^{(1,2)}$ and $S_{\AL}^{(2,1)}$ given by Eqs.~(\ref{eq:SAL12}) and (\ref{eq:SAL21}), respectively. Then, using the FDT, we find the first two leading terms in the real part of the AL conductivity to be given by Eq.~(\ref{eq:sigmaAL}).


\end{document}